\documentclass[twocolumn,showpacs,aps,epsfig]{revtex4}

\usepackage{graphicx}
\usepackage{epstopdf}
\usepackage{latexsym}

\usepackage[center]{subfigure}

\begin{document}

 \newcommand{\bq}{\begin{equation}}
 \newcommand{\eq}{\end{equation}}
 \newcommand{\bqn}{\begin{eqnarray}}
 \newcommand{\eqn}{\end{eqnarray}}
 \newcommand{\nb}{\nonumber}
 \newcommand{\lb}{\label}
\newcommand{\PRL}{Phys. Rev. Lett.}
\newcommand{\PL}{Phys. Lett.}
\newcommand{\PR}{Phys. Rev.}
\newcommand{\CQG}{Class. Quantum Grav.}

\title{The cosmological constant in the brane world of string theory on $S^{1}/Z_{2}$}
\author{  Anzhong Wang ${}^{1,2,3}$ and N.O. Santos  ${}^{3, 4, 5}$} 
\affiliation{ ${}^{1} $ GCAP-CASPER, Department of Physics, Baylor University,
Waco, Texas 76798-7316\\
${}^{2}$ Department of Theoretical Physics, the State University of
Rio de Janeiro, RJ, Brazil\\
${}^{3} $ LERMA/CNRS-FRE 2460, Universit\'e Pierre et Marie Curie, ERGA, 
Bo\^{\i}te 142, 4 Place Jussieu, 75005 Paris Cedex 05, France\\
${}^{4} $  School of Mathematical Sciences, Queen Mary,
University of London, London E1 4NS, UK\\ 
${}^{5}$ Laborat\'orio Nacional de Computa\c{c}\~{a}o Cient\'{\i}fica, 
25651-070 Petr\'opolis RJ, Brazil}
 
\date{\today}

\begin{abstract}

Orbifold branes in string theory are investigated, and the general field
equations both outside and on the branes are given explicitly for type II 
and heterotic string. The radion stability is studied using the 
Goldberger-Wise mechanism, and shown explicitly that it is stable. 
It is also found that the effective cosmological constant on  each of 
the two branes can be easily lowered to its current observational value, 
using large extra dimensions. This is also true for type I string. 
Therefore, brane world of string theory provides a viable and built-in 
mechanism for solving the long-standing cosmological constant problem. 
Applying the formulas to cosmology, we obtain the generalized Friedmann 
equations on the branes. 

\end{abstract}


\maketitle

\vspace{1.cm}

\section{Introduction}

\renewcommand{\theequation}{1.\arabic{equation}}
\setcounter{equation}{0}


One of the long-standing problems is the cosmological constant 
(CC) problem: its theoretical expectation values  from quantum field theory
exceed observational limits by $120$ orders of magnitude  \cite{wen}. Even if
such high energies are suppressed by supersymmetry, the electroweak corrections  
are still $56$ orders higher. This problem was further sharpened by recent  
observations of supernova (SN) Ia, which  reveal the striking discovery that our 
universe has lately been in its accelerated expansion phase  \cite{agr98}. Cross 
checks from the cosmic microwave background radiation  and large scale structure 
all confirm it \cite{Obs}. In Einstein's theory of gravity, such an expansion can 
be achieved by a tiny positive CC, which is well 
consistent with all observations carried out so far \cite{SCH07}. Because of this
remarkable fact, a large number of ambitious projects have been aimed to distinguish 
the CC from dynamical models \cite{DETF}. 

Therefore,  solving the CC problem now becomes more urgent than 
ever before. Since the problem is intimately related to 
quantum gravity, its solution is expected to come from quantum gravity, too. At 
the present, string/M-Theory is our best bet for a consistent  quantum theory of 
gravity, so it is  reasonable to ask what string/M-Theory has to say about the CC. 
In the string landscape \cite{Susk}, it is expected there are many different vacua 
with different local CC's \cite{BP00}. Using the anthropic principle, 
one may select the low energy vacuum in which we can exist. However, many theorists 
still hope to explain the problem without invoking the existence of ourselves.  
 
Lately,  the CC problem and the late transient  
acceleration  of the universe was studied  \cite{GWW07}  in the framework of the 
Horava-Witten heterotic M-Theory on $S^{1}/Z_{2}$ \cite{HW96}. Using the 
Arkani-Hamed-Dimopoulos-Dvali (ADD) mechanism of large extra dimensions \cite{ADD98}, 
it was shown that the effective CC on each of the two branes can be easily 
lowered to its current observational value. The domination of this term  
is only temporary.  Due to the interaction of the bulk and the brane, the universe 
will be in its decelerating expansion phase again, whereby all problems connected 
with a far future de Sitter universe 
\cite{KS00} are resolved.

In this Letter, we shall generalize the above studies to string theory, and show that
the same mechanism is also viable in all of the five versions of string theory. In
addition, we also study the radion stability, using the Goldberger-Wise mechanism
\cite{GW99}, and show explicitly that the radion is indeed stable in our current setup. 
Thus, brane world of string/M-Theory provides  a built-in mechanism  for solving 
the long-standing CC problem. 

Before showing our above claims, we note that   the CC problem was also
studied in the framework of brane world in 5D spacetimes \cite{5CC} and 6D 
supergravity \cite{6CC}. However, it turned out that in the 5D case hidden 
fine-tunings are required \cite{For00}, while in the 6D case it is still not clear 
whether loop corrections can be as small as  required \cite{Burg07}.

\section{The Model}

\renewcommand{\theequation}{2.\arabic{equation}}
\setcounter{equation}{0}

Let us begin with the toroidal compactification of the Neveu-Schwarz/Neveu-Schwarz 
(NS-NS) sector of the action in (D+d) dimensions, $\hat{M}_{D+d} = M_{D}\times {\cal{T}}_{d}$, 
where  ${\cal{T}}_{d}$ is a d-dimensional torus.  The action takes the form 
\cite{LWC00},
\bqn
\lb{2.1}
\hat{S}_{D+d} &=& -\frac{1}{2\kappa^{2}_{D+d}}
\int_{M^{D+d}}{\sqrt{\left|\hat{g}_{D+d}\right|}  e^{-\hat{\Phi}} \left\{
{\hat{R}}_{D+d}[\hat{g}]\right.}\nb\\
& & \left. + \hat{g}^{AB}\left(\hat{\nabla}_{A}\hat{\Phi}\right)
\left(\hat{\nabla}_{B}\hat{\Phi}\right) - \frac{1}{12}{\hat{H}}^{2}\right\},
\eqn
where $\hat{\nabla}_{A}$ denotes the covariant derivative with respect to $\hat{g}^{AB}$
with $A, B = 0, 1, ..., D+d-1$; $\hat{\Phi}$ is the dilaton field; $\hat{H} \equiv dB$ 
describes the NS
three-form field strength of the fundamental string; and $\kappa^{2}_{D+d}$ is the gravitational
coupling constant. 
It should be noted that such action is common to both
type II and heterotic string \cite{LWC00}. For type I string, the dilaton does not
couple conformally with the NS three-form. However, as to be shown below, our conclusions 
can be easily generalized to the latter  case. In particular, our results about the CC
 are equally applicable to type I string.

The $(D+d)$-dimensional spacetimes are described by the metric,
\bq
\lb{2.3}
d{\hat{s}}^{2}_{D+d} 
= \tilde{g}_{ab}\left(x^{c}\right) dx^{a}dx^{b}  +   h_{ij}\left(x^{c}\right)dz^{i} dz^{j},
\eq
where  $\tilde{g}_{ab}$ is the metric on $M_{D}$, parametrized by the coordinates $x^{a}$
with $a,b, c = 0, 1, ..., D-1$, and $h_{ij}$ is the metric on the compact space ${\cal{T}}_{d}$
with the periodic coordinates $z^{i}$, where $i, j = D, D+1, ..., D+d-1$. 
We assume that all the matter fields are functions of $x^{a}$ only. This implies that the 
compact space  ${\cal{T}}_{d}$ is Ricci flat, $R_{d}[h] = 0$.
Moreover, following  \cite{BW06}   we also add a potential term to the total action,
$\hat{S}^{potential}_{D+d} = \int_{M^{D+d}}{\sqrt{\left|\hat{g}_{D+d}\right|} V_{D+d}^{s}}$.
Then, after the dimensional reduction, the D-dimensional action in the Einstein frame takes
the form,
\bq
\lb{2.16}
S_{D}^{(E)} = -
\int_{M^{D}}{\sqrt{\left|{g}_{D}\right|}  
\left\{\frac{1}{2\kappa^{2}_{D}}{R}_{D}[{g}] - {\cal{L}}_{D}^{(E)}(\phi, \psi, B)\right\}},
\eq
where  
\bqn
\lb{2.16a}
& &  {\cal{L}}_{D}^{(E)} \equiv \frac{1}{12}\left\{ 6\left[\left(\nabla\phi\right)^{2} 
+ \left(\nabla\psi\right)^{2} - 2V_{D}\right]  + 3 e^{- \sqrt{\frac{8}{d}}\; \psi} \right.\nb\\
& & \;\; \left. \times \left({\nabla}_{a}B^{ij}\right)
\left({\nabla}^{a}B_{ij}\right)  +  e^{- \sqrt{\frac{8}{D-2}}\; \phi} H_{abc}H^{abc}\right\},  
\eqn
where $B^{ij} \equiv \delta^{ik}\delta^{jl} B_{kl}$, 
and $\nabla_{a}$ denotes the covariant derivative with respect to $g_{ab}$, which is 
related to the string metric $\tilde{g}_{ab}$ by
$g_{ab} = \Omega^{2}\tilde{g}_{ab}$, where $
\Omega^{2} = \exp\left(-{2}\tilde{\phi}/{(D-2)}\right)$, 
 $\phi = \sqrt{{2}/{(D-2)}} \; \tilde{\phi}$, and
$\tilde{\phi} = \hat{\Phi} - (1/2)\ln\left|h\right|$.   
$\kappa^{2}_{D}$ is defined as $\kappa^{2}_{D} \equiv 
V_{0}^{-1}\kappa^{2}_{D+d}$, with    
$V_{0} \equiv \int{d^{d}z}$. Note that in writing down the above action, we assumed that (a) the
flux $B$ is block diagonal; and (b) the internal metric takes the form,
$h_{ij} = - \exp\left(\sqrt{2/d}\; \psi\right)\delta_{ij}$. Then,   we find that 
$V_{D} = V^{0}_{D}   \exp\left({D}/{\sqrt{2(D-2)}}\;\phi 
 + \sqrt{{d}/{2}}\;\psi\right)$, where $V^{0}_{D} \equiv 2\kappa^{2}_{D}V_{0} V_{D+d}^{s}$.

To study orbifold branes, we  add the brane actions,
\bqn
\lb{3.1}
S^{(I)}_{D-1, m} &=&  - \int_{M^{(I)}_{D-1}}{\sqrt{\left|g^{(I)}_{D-1}\right|}
\left(\tau^{(I)}_{(\phi,\psi)} + g^{(I)}_{k}\right. }\nb\\
& & \left. - {\cal{L}}^{(I)}_{D-1,m}\left(\phi, \psi, B, \chi\right)\right),
\eqn
where, $I = 1, 2$, and $\tau^{(I)}_{(\phi,\psi)} \equiv \epsilon_{I}V^{(I)}_{D-1}(\phi,\psi)$,
with $V^{(I)}_{D-1}(\phi, \psi)$  
denoting the potential of the scalar fields,  and $\epsilon_{1} = - \epsilon_{2} = 1$.   
$\chi$  denotes collectively all matter fields, and $ g^{(I)}_{\kappa}$ are constants, 
as to be shown below,  directly related to the $(D-1)$-dimensional Newtonian constant 
$G_{D-1}^{(I)}$.  Then,   the   field equations  take the form, 
\bq
\lb{3.9}
G^{(D)}_{ab} = \kappa^{2}_{D} T^{(D)}_{ab} + \kappa^{2}_{D}  
\sum^{2}_{i=1}{{\cal{T}}^{(I)}_{ab}  
 \sqrt{\left|\frac{g^{(I)}_{D-1}}{g_{D}}\right|}\; \delta\left(\Phi_{I}\right)},
\eq
where $\kappa^{2}_{D}T^{(D)}_{ab} \equiv 2\delta{\cal{L}}_{D}^{(E)}/
\delta{g^{ab}} -   g_{ab}{\cal{L}}_{D}^{(E)}$;
${\cal{T}}^{(I)}_{ab} \equiv 
{\cal{T}}^{(I)}_{\mu\nu} e^{(I, \; \mu)}_{a}e^{(I, \; \nu)}_{b}$;
$ {\cal{T}}^{(I)}_{\mu\nu} =  
{\tau}^{(I)}_{\mu\nu} 
+  \left(\tau^{(I)}_{(\phi,\psi)} + g^{(I)}_{k}\right) g_{\mu\nu}^{(I)}$;
${\tau}^{(I)}_{\mu\nu} \equiv 2 {\delta{\cal{L}}^{(I)}_{D-1,m}}/
{\delta{g^{(I)\; \mu\nu}}}
-   g^{(I)}_{\mu\nu}{\cal{L}}^{(I)}_{D-1,m}$;
and $e^{(I)\; a}_{(\mu)} \equiv \partial x^{a}/\partial \xi^{\mu}_{(I)}$.   
$\xi_{(I)}^{\mu}$ are the intrinsic coordinates of the I-th brane with $\mu, 
\nu = 0, 1, 2, ..., D-2$;
$g_{\mu\nu}^{(I)}$  is the reduced metric on the I-th brane, 
$g_{\mu\nu}^{(I)} \equiv \left. e^{(I)a}_{(\mu)} e^{(I)b}_{(\nu)} 
g_{ab}\right|_{M^{(I)}_{D-1}}$;
$\Phi_{I}\left(x^{a}\right)  = 0$ denotes the location of the I-th brane;
and $\delta(x)$  the Dirac delta function.


To write down the field equations on the branes, 
we use the Gauss-Codacci equations  \cite{SMS},
 \bqn
 \lb{3.12}
 G^{(D-1)}_{\mu\nu} &=& {\cal{G}}^{(D)}_{\mu\nu} + E^{(D)}_{\mu\nu}
 + {\cal{F}}^{(D-1)}_{\mu\nu},\\
 \lb{3.13}
 {\cal{F}}^{(D-1)}_{\mu\nu} &\equiv&  
 K_{\mu\lambda}K^{\lambda}_{\nu} - KK_{\mu\nu}  
  - \frac{1}{2}g_{\mu\nu}\left(K_{\alpha\beta}K^{\alpha\beta} 
    - K^{2}\right),\nb\\
 {\cal{G}}_{\mu\nu}^{(D)} &\equiv&  \frac{(D-3)}{(D-2)(D-1)}
\left\{(D-1)G_{ab}^{(D)}e^{a}_{(\mu)} e^{b}_{(\nu)} \right.\nb\\
& & \left.
- \left[(D-1)G_{ab}n^{a}n^{b} + G^{(D)}\right]g_{\mu\nu}\right\},  
\eqn
where $n^{a}$ denotes the normal vector to the brane, $G^{(D)}
\equiv g^{ab} G^{(D)}_{ab}$, $E^{(D)}_{\mu\nu} 
\equiv C_{abcd}^{(D)}n^{a}e^{b}_{(\mu)}n^{c}e^{d}_{(\nu)}$,
and $C_{abcd}^{(D)}$ is the D-dimensional Weyl tensor.  The extrinsic curvature 
$K_{\mu\nu}$  is defined as
$K_{\mu\nu} \equiv e^{a}_{(\mu)}e^{b}_{(\nu)}\nabla_{a}n_{b}$.

Assuming that the branes have $Z_{2}$ symmetry, we can express the intrinsic
curvatures $K^{(I)}_{\mu\nu}$ in terms of the effective energy-momentum tensor
${\cal{T}}_{\mu\nu} ^{(I)}$ through the Lanczos equations  \cite{Lan22},
$\left[K_{\mu\nu}^{(I)}\right]^{-} - g_{\mu\nu}^{(I)} \left[K^{(I)}\right]^{-} 
= - \kappa^{2}_{D}{\cal{T}}_{\mu\nu} ^{(I)}$,
where $\left[K_{\mu\nu}^{(I)}\right]^{-} \equiv {\rm lim}_{\Phi_{I} \rightarrow 0^{+}}
K^{(I)\; +}_{\mu\nu} - {\rm lim}_{\Phi_{I} \rightarrow 0^{-}}
K^{(I)\; -}_{\mu\nu}$, and $\left[K^{(I)}\right]^{-} \equiv g^{(I)\; \mu\nu}
\left[K_{\mu\nu}^{(I)}\right]^{-}$. 
Setting $ {\cal{S}}^{(I)}_{\mu\nu} = \tau^{(I)}_{\mu\nu} + \lambda^{(I)}g^{(I)}_{\mu\nu}$,
 where $\lambda^{(I)}$ denotes the CC of the I-th brane, we find that
 $ G^{(D-1)}_{\mu\nu}$ given by Eq.(\ref{3.12}) can be cast in the form,
 \bqn
 \lb{3.15}
 G^{(D-1)}_{\mu\nu} &=& {\cal{G}}^{(D)}_{\mu\nu} + E^{(D)}_{\mu\nu}
 + {\cal{E}}_{\mu\nu}^{(D-1)} + \kappa^{4}_{D}\pi_{\mu\nu}\nb\\
 & & + \kappa^{2}_{D-1}\tau_{\mu\nu} + \Lambda_{D-1} g_{\mu\nu},\\
\lb{3.16}
\pi_{\mu\nu} &\equiv& \frac{1}{8(D-2)}\left\{2(D-2)\tau_{\mu\lambda}\tau^{\lambda}_{\nu}
-  2\tau \tau_{\mu\nu}\right.\nb\\
& & \left. 
 -  g_{\mu\nu}\left((D-2)\tau^{\alpha\beta} \tau_{\alpha\beta}
 -  \tau^{2}\right)\right\},\\
 \lb{3.16a}
 {\cal{E}}_{\mu\nu}^{(D-1)} &\equiv& \frac{\kappa^{4}_{D}(D-3)}{8(D-2)}\tau_{(\phi,\psi)}\nb\\
 & & \times \left[2\tau_{\mu\nu}  
   + \left(2\lambda + \tau_{(\phi,\psi)}\right)g_{\mu\nu}\right],\\
\lb{3.17}
\frac{\kappa^{2}_{D-1}}{\kappa^{4}_{D}} &=& \frac{(D-3)\lambda}{4(D-2)},\;\;
\frac{\Lambda_{D-1}}{\kappa^{4}_{D}}  =  \frac{(D-3)\lambda^{2}}{8(D-2)}.
\eqn
Note that in writing Eqs.(\ref{3.15})-(\ref{3.17}), without causing any confusion, 
we had dropped the super indices $``(I)"$. In addition, the definitions of  
$\kappa_{D-1}$ and  $\Lambda$ are unique, because these are the only terms  that linearly 
proportional to  the matter field $\tau_{\mu\nu}$ and the spacetime geometry $g_{\mu\nu}$. 
When $D = 5$ they reduce exactly to the ones defined in brane-worlds \cite{branes}.
 
 In the following, we shall restrict ourselves to the case where $D = d = 5$.

\section{Radion Stability}

\renewcommand{\theequation}{3.\arabic{equation}}
\setcounter{equation}{0}

\begin{figure}
\includegraphics[width=\columnwidth]{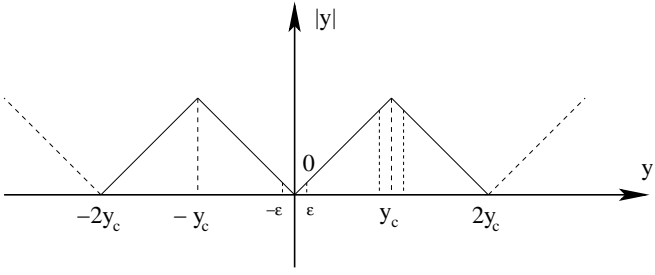}
\caption{The function $\left|y\right|$  appearing in the metric Eq.(\ref{4.4a}). }
\label{fig2}
\end{figure}

 In the studies of orbifold branes, an important issue is the  radion stability.
In this section, we shall address this problem. Let us first consider
the 5-dimensional static metric with a 4-dimensional Poincar\'e symmetry,
\bqn
\lb{5.1a}
ds^{2}_{5} &=& e^{2\sigma(y)}\left(\eta_{\mu\nu}dx^{\mu}dx^{\nu} - dy^{2}\right),\\
\lb{5.1b}
\sigma(y) &=&  \frac{1}{9}\ln\left(\frac{|y| + y_{0}}{L}\right), \nb\\
\phi(y) &=& - \sqrt{\frac{25}{54}}\; \ln\left(\frac{|y| + y_{0}}{L}\right) + \phi_{0},\nb\\
\psi(y) &=& -  \sqrt{\frac{5}{18}}\; \ln\left(\frac{|y| + y_{0}}{L}\right)
+ \psi_{0}, \nb\\
B_{ij} &=& 0 = B_{ab},
\eqn
where $|y|$ is defined as that given in Fig.\ref{fig2} \cite{WCS08}, 
${L}$ and $y_{0}$ are positive constants, and
\bq
\lb{5.1c}
\psi_{0} \equiv \sqrt{\frac{2}{5}}\left(\ln\left(\frac{2}{9{L}^{2}V^{0}_{(5)}}\right)
- \frac{5}{\sqrt{6}}\phi_{0}\right).
\eq

Then, it can be shown that  the above solution satisfies the  gravitational and matter 
field equations outside the branes, Eq.(\ref{3.9}), for $D = d = 5$. On the other hand, to 
show that it also satisfies the field equations on the branes, we first note that
the normal vector $n^{a}_{(I)}$  to the I-th brane is given simply by
\bq
\lb{5.1d}
n^{a}_{(I)} = - \epsilon_{y}^{(I)} e^{-\sigma(y_{I})}\delta^{a}_{y},
\eq
and that
\bq
\lb{5.1e}
\dot{t} = e^{-\sigma(y_{I})}, 
\;\; \dot{y} = 0,
\eq 
where $ y_{1} = y_{c}> 0$ and $y_{2} = 0$. 
Inserting the above into Eqs.(\ref{3.15})-(\ref{3.17}),   we find that they are satisfied for 
$\tau^{(I)}_{\mu\nu} = 0$, provided that the tension $\tau_{\phi, \psi}^{(I)}$
  satisfies the relation,
\bq
\lb{5.2}
\left(\tau^{(I)}_{(\phi, \psi)} + 2\rho^{(I)}_{\Lambda}\right)^{2}
= \frac{\rho^{(I)}_{\Lambda}}{54\pi G_{4} L^{2}} \left(\frac{{L}}{y_{I} + y_{0}}\right)^{20/9},
\eq
where $\rho^{(I)}_{\Lambda} \equiv \Lambda^{(I)}/(8\pi G_{4})$ denotes the corresponding  
energy density of the 
effective cosmological constant on the I-th brane, defined by Eq.(\ref{3.17}).
On the other hand, on each of the two branes, we also find that
\bqn
\lb{5.2ac}
\frac{\partial{V^{(I)}_{4}}(\phi, \psi)}{\partial{\phi}} &=& 
\sqrt{\frac{25}{54}} \; \frac{\epsilon_{y}^{(I)}}{\kappa_{5}^{2}\left(y_{I} + y_{0}\right)},\\
\lb{5.2ad}
\frac{\partial{V^{(I)}_{4}}(\phi, \psi)}{\partial{\psi}} &=& 
\sqrt{\frac{5}{18}} \; \frac{\epsilon_{y}^{(I)}}{\kappa_{5}^{2}\left(y_{I} + y_{0}\right)}.
\eqn
For certain choices of the potentials $V^{(I)}_{4}(\phi, \psi)$ of the two branes, 
Eqs.(\ref{5.2})-(\ref{5.2ad}) can be satisfied. For example, one may choose
\bq
\lb{5.2ae}
 V^{(I)}_{4}(\phi, \psi)  = \beta_{I} 
 e^{-g_{I}\phi}\left(\psi^{2} - \psi_{I}^{2}\right)^{2},  
\eq
where $\beta_{I}, \; g_{I}$ and $\psi_{I}$ are arbitrary constants.
Then, by properly choosing these parameters,  Eqs.(\ref{5.2})-(\ref{5.2ad})
can easily be satisfied.

To study  the radion stability,
it is found convenient to introduce the proper distance $Y$, defined by
\bq
\lb{5.2a}
Y = \left(\frac{9L}{10}\right)\left\{\left(\frac{y+y_{0}}{L}\right)^{10/9} 
- \left(\frac{y_{0}}{L}\right)^{10/9}\right\}.
\eq
Then, in terms of $Y$, the static solution (\ref{5.1a}) can be written as
\bq
\lb{5.1aa}
ds^{2}_{5} = e^{-2A(Y)}\eta_{\mu\nu}dx^{\mu}dx^{\nu} - dY^{2},
\eq
with 
\bqn
\lb{5.1ab}
A(Y) &=&  -\frac{1}{10}\ln\left\{\left(\frac{10}{9L}\right) \left(|Y| + Y_{0}\right)\right\},\nb\\
\phi(Y) &=&  -\sqrt{\frac{3}{8}}\ln\left\{\left(\frac{10}{9L}\right) \left(|Y| + Y_{0}\right)\right\}
+ \phi_{0},\nb\\
\psi(Y) &=&  -\frac{3}{\sqrt{40}}\ln\left\{\left(\frac{10}{9L}\right) \left(|Y| + Y_{0}\right)\right\}\nb\\
& & + \psi_{0},
\eqn
where $|Y|$ is defined  also as that in Fig. \ref{fig2},
with  
\bqn
\lb{5.2b}
Y_{c} &\equiv& \left(\frac{9L}{10}\right)\left\{\left(\frac{y_{c} + y_{0}}{L}\right)^{10/9} 
- \left(\frac{y_{0}}{L}\right)^{10/9}\right\},\nb\\
Y_{0} &\equiv& \left(\frac{9L}{10}\right)\left(\frac{y_{0}}{L}\right)^{10/9},
\eqn
and $Y_{2} = 0, \; Y_{1} = Y_{c}$.

Following \cite{GW99}, let us consider a massive scalar field $\Phi$ with the actions,
\bqn
\lb{5.3a}
S_{b} &=& \int{d^{4}x}\int_{0}^{Y_{c}}{dY \sqrt{-g_{5}} 
\left(\left(\nabla\Phi\right)^{2} - m^{2}\Phi^{2}\right)},\nb\\
S_{I} &=&  - \alpha_{I} \int_{M^{(I)}_{4}}{d^{4}x \sqrt{-g_{4}^{(I)}} 
\left(\Phi^{2} - v^{2}_{I}\right)^{2}},
\eqn
where $\alpha_{I}$ and $v_{I}$ are real constants.  Then,  it can be shown 
that, in the background of Eq.(\ref{5.1aa}),  the massive scalar field $\Phi $ satisfies
the following Klein-Gordon equation
\bq
\lb{5.5}
\Phi'' - 4A'\Phi' - m^{2}\Phi =
\sum_{I=1}^{2}{2  \alpha_{I} \Phi \left(\Phi^{2} - v^{2}_{I}\right)\delta(Y - Y_{I})}.
\eq
Integrating the above equation in the neighborhood of the I-th brane, we find that
\bq
\lb{5.5a}
\left. \frac{d\Phi(Y)}{dY}\right|_{Y_{I}-\epsilon}^{Y_{I}+\epsilon}
= 2\alpha_{I}  \Phi_{I} \left(\Phi^{2}_{I} - v^{2}_{I}\right),
\eq
where $\Phi_{I} \equiv \Phi(Y_{I})$. Setting
\bq
\lb{5.5b}
z \equiv m\left(Y + Y_{0}\right),\;\;\;
\Phi = z^{\nu}\; u(z),
\eq
we find that, outside of the branes, Eq.(\ref{5.5}) yields,
\bq
\lb{5.5c}
\frac{d^{2}u}{dz^{2}} + \frac{1}{z}\frac{du}{dz}
- \left(1 + \frac{\nu^{2}}{z^{2}}\right)u = 0,
\eq
where $\nu \equiv 3/10$. Eq.(\ref{5.5c}) is the standard modified Bessel equation
\cite{AS72}, which has the general solution
\bq
\lb{5.5d}
u(z) = a I_{\nu}(z) + b K_{\nu}(z),
\eq
where $I_{\nu}(z)$ and $K_{\nu}(z)$ denote the modified
Bessel functions, and  $a$ and $b$ are the integration constants, which are uniquely 
determined by the boundary conditions (\ref{5.5a}). Since
\bqn
\lb{5.5e}
\lim_{Y \rightarrow Y_{c}^{+}}{\frac{d\Phi(Y)}{dY}} &=& 
- \lim_{Y \rightarrow Y_{c}^{-}}{\frac{d\Phi(Y)}{dY}}   
\equiv - \Phi'\left(Y_{c}\right),\nb\\
\lim_{Y \rightarrow 0^{-}}{\frac{d\Phi(Y)}{dY}} &=& 
- \lim_{Y \rightarrow 0^{+}}{\frac{d\Phi(Y)}{dY}} \equiv  -\Phi'(0),
\eqn
we find that the  conditions  (\ref{5.5a}) can be written in the forms,
\bqn
\lb{5.5fa}
\Phi'(Y_{c}) &=& - \alpha_{1}  \Phi_{1} \left(\Phi^{2}_{1} - v^{2}_{1}\right),\\
\lb{5.5fb}
\Phi'(0) &=& \alpha_{2}  \Phi_{2} \left(\Phi^{2}_{2} - v^{2}_{2}\right).
\eqn

Inserting the above solution back to the actions (\ref{5.3a}), and then integrating
them with respect to $Y$, we obtain the effective potential for the radion $Y_{c}$,
\bqn
\lb{5.5g}
V_{\Phi}\left(Y_{c}\right) &\equiv&  - \int_{0+\epsilon}^{Y_{c}- \epsilon}{dY \sqrt{\left|g_{5}\right|} 
\left(\left(\nabla\Phi\right)^{2} - m^{2}\Phi^{2}\right)}\nb\\
& & + \sum_{I=1}^{2}{ \alpha_{I}  \int_{Y_{I} -\epsilon}^{Y_{I} + \epsilon}
{dY \sqrt{\left|g_{4}^{(I)}\right|} 
\left(\Phi^{2} - v^{2}_{I}\right)^{2}}}\nb\\
& & \;\;\;\;\;\;\; \;\;\;\;\;\;\; \times \delta\left(Y-Y_{I}\right) \nb\\
&=& \left. e^{-4A(Y)}\Phi(Y)\Phi'(Y)\right|^{Y_{c}}_{0}\nb\\
& &
+ \sum_{I=1}^{2}{\alpha_{I} \left(\Phi^{2}_{I} - v^{2}_{I}\right)^{2}e^{-4A(Y_{I})}}.
\eqn	 
In the limit that $\alpha_{I}$'s are very large \cite{GW99}, Eqs.(\ref{5.5fa}) 
and (\ref{5.5fb}) show that there are solutions only when
$\Phi(0) = v_{2}$ and $\Phi(Y_{c}) = v_{1}$, that is,
\bqn
\lb{5.5ha}
v_{1} &=& z_{c}^{\nu}\left(a I_{\nu}\left(z_{c}\right) +
b K_{\nu}\left(z_{c}\right)\right),\\
\lb{5.5hb}
v_{2} &=&  z_{0}^{\nu}\left(a I_{\nu}\left(z_{0}\right) +
b K_{\nu}\left(z_{0}\right)\right),
\eqn
where $z_{0} \equiv   mY_{0}$ and $z_{c} \equiv  m(Y_{c} + Y_{0})$. Eqs.(\ref{5.5ha})
and (\ref{5.5hb}) have the solutions,
\bqn
\lb{5.5i}
a &=& \frac{1}{\Delta}\left(K^{(0)}_{\nu} z^{\nu}_{0} v_{1}
- K^{(c)}_{\nu} z^{\nu}_{c}v_{2}\right),\nb\\
b &=& \frac{1}{\Delta}\left(I^{(c)}_{\nu} z^{\nu}_{c} v_{2}
- I^{(0)}_{\nu} z^{\nu}_{0} v_{1}\right),
\eqn
where $K^{(I)}_{\nu} \equiv K_{\nu}(z_{I}),\;
I^{(I)}_{\nu} \equiv I_{\nu}(z_{I})$, and
\bq
\lb{5.5j}
\Delta \equiv \left(z_{0}z_{c}\right)^{\nu}
\left(I^{(c)}_{\nu} K^{(0)}_{\nu} - I^{(0)}_{\nu} K^{(c)}_{\nu}\right).
\eq

\subsection{$ m Y_{0} \gg 1$}
 
\begin{figure}
\centering
\includegraphics[width=8cm]{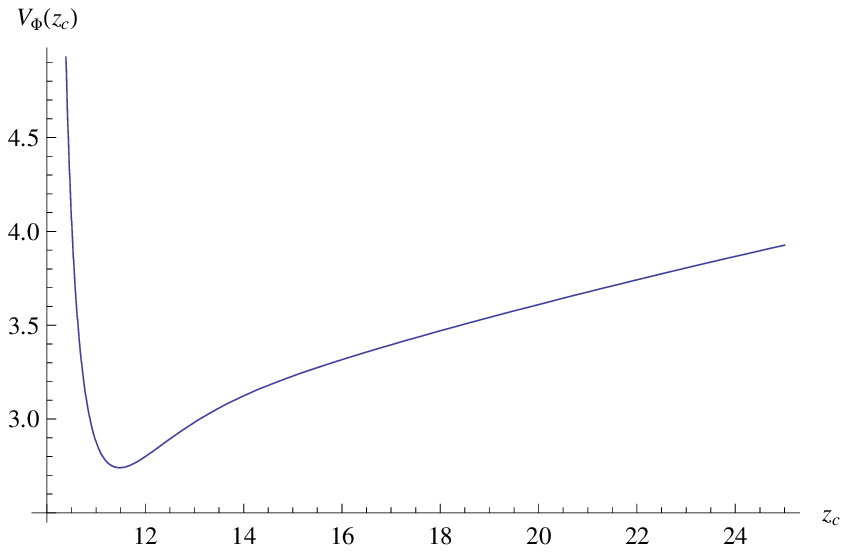}
\caption{The  potential defined by Eq.(\ref{5.4}) in the  limit of large $v_{I}$ and $y_{0}$.
In this particular plot, we choose $(z_{0}, \;  v_{1}, \;  v_{2}) = (10, \;  1.0,\; 0.1)$. }
\label{fig3}
\end{figure}

\begin{figure}
\centering
\includegraphics[width=8cm]{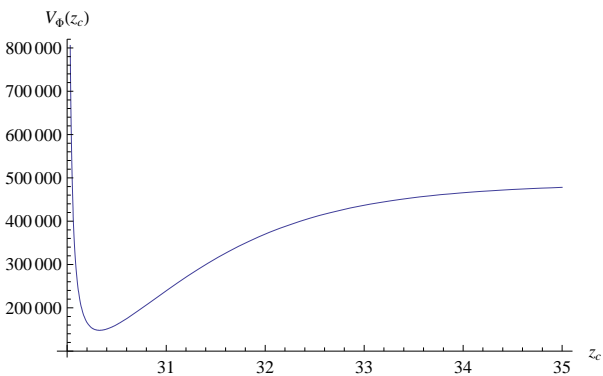}
\caption{The  potential defined by Eq.(\ref{5.4}) in the  limit of large $v_{I}$ and $y_{0}$.
In this particular plot, we choose $(z_{0}, \;  v_{1}, \;  v_{2}) = (30, \;  200,\; 100)$. }
\label{fig4}
\end{figure}

When $Y_{0} \gg m^{-1}$, we have $z_{0}, \; z \gg 1$. Then, we find that \cite{AS72},
\bqn
\lb{5.5k}
I_{\nu}(z) &\simeq& \frac{e^{z}}{\sqrt{2\pi z}}  \simeq I_{\nu}'(z),\nb\\
K_{\nu}(z) &\simeq& \sqrt{\frac{\pi}{2z}} e^{-z}  \simeq - K_{\nu}'(z).
\eqn
Substituting them into Eq.(\ref{5.5g}), we find that  
\bqn
\lb{5.4}
V_{\Phi}\left(Y_{c}\right) &=& \frac{1}{2} m^{3/5}\left(\frac{10}{9}\right)^{2/5}
\frac{e^{-\left(z_{0} + z_{c}\right)}}{\left(z_{0}z_{c}\right)^{3/5}\sinh\left(z_{c}-z_{0}\right)}\nb\\
& & \times \left\{2\nu e^{z_{0} + z_{c}} \sinh\left(z_{c}-z_{0}\right)
\left(v^{2}_{1}z^{3/5}_{0} - v^{2}_{2}z^{3/5}_{c}\right) \right.\nb\\
& & + \left(z_{0}z_{c}\right)^{3/5}
   \left[\left(v_{1}z^{1/5}_{c}e^{z_{0}} - v_{2}z^{1/5}_{0}e^{z_{c}}\right)^{2}\right.\nb\\
& &   \left.\left. 
    + \left(v_{1}z^{1/5}_{c}e^{z_{c}} - v_{2}z^{1/5}_{0}e^{z_{0}}\right)^{2}\right]\right\}.
\eqn
Then, we find that
\bq
\lb{5.6}
V_{\Phi}\left(Y_{c}\right) = V_{\Phi}^{(0)}\left\{\matrix{\frac{(v_{1} - v_{2})^{2}z^{2/5}_{0}}
{\sinh\left(z_{c}-z_{0}\right)} \rightarrow \infty, & z_{c} \rightarrow z_{0},\cr 
 v_{1}^{2}  z_{c}^{2/5} \rightarrow \infty, & z_{c} \rightarrow \infty,\cr}\right.
\eq
where $V_{\Phi}^{(0)} \equiv m^{3/5}\left(\frac{10}{9}\right)^{2/5}$.
Figs. \ref{fig3} and \ref{fig4} show the potential for $(z_{0}, \;  v_{1}, \;  v_{2}) 
= (10, \;  1.0,\;0.1)$ and $(z_{0}, \;  v_{1}, \;  v_{2}) = (30, \;  200,\;100)$, 
respectively, from which we can see clearly that it has a minimum.  Therefore, the radion is 
indeed stable in our current setup.

\subsection{$ m Y_{0} \ll 1$}

When $ m Y_{0} \ll 1$ and $ m Y_{c} \ll 1$, we find that \cite{AS72}
\bqn
\lb{5.5ka}
I_{\nu}(z) &\simeq& \frac{z^{\nu}}{2^{\nu} \Gamma(\nu + 1)},\nb\\
K_{\nu}(z) &\simeq& \frac{2^{\nu - 1} \Gamma(\nu)}{z^{\nu}}.
\eqn
Substituting them into Eq.(\ref{5.5g}), we obtain  
\bq
\lb{5.4a}
V_{\Phi}\left(Y_{c}\right) = \frac{3}{5} m^{3/5}\left(\frac{10}{9}\right)^{2/5}
\frac{\left(v_{1} - v_{2}\right)^{2}}{z_{c}^{2\nu} - z_{0}^{2\nu}}.
\eq
Clearly, in this limit the potential has no minima, and the corresponding radion is not stable. 
Therefore, there exists a minimal mass for the scalar field $\Phi$, say, $m_{c}$, only when 
$m > m_{c}$ the corresponding radion is stable. 
 
It should be noted that, in the Randall-Sundrum setup \cite{RS1}, $Y_{c}$ is required to be
about $35$ in order to solve the hierarchy problem. However, in the current setup the hierarchy 
problem is solved by the combination of the RS warped factor mechanism and the ADD large
extra dimensions  \cite{WS08}. Thus, such a requirement is not needed here, which  allows 
$Y_{c}$ to have a large range of choice.

\section{The Cosmological Constant}

\renewcommand{\theequation}{4.\arabic{equation}}
\setcounter{equation}{0}

For $D = d  =5$, we find that 
\bq
\lb{3.18a}
\kappa^{2}_{5} = \frac{\kappa^{2}_{10}}{V_{0}} = \frac{1}{M^{8}_{10}R^{5}}.
\eq
Then, from Eq.(\ref{3.17}) we find
\bq
\lb{3.18}
\rho_{\Lambda} \equiv \frac{\Lambda_{4}}{8\pi G_{4}} = 
3 \left(\frac{R}{{{l}}_{pl}}\right)^{10}\left(\frac{M_{10}}{M_{pl}}\right)^{16}
{M_{pl}}^{4},
\eq
where $R$ denotes the typical size of the internal space ${\cal{T}}_{d}$, and ${M_{pl}}$ and 
${{l}}_{pl}$ denote the Planck mass and length,  respectively. Current observations show
 $\rho_{\Lambda} \simeq 10^{-47}\; GeV^{4}$. If $M_{10}$ is of the order of TeV  \cite{Anto00}, 
we find that Eq.(\ref{3.18}) requires $R \simeq 10^{-22} \; m$, which is well below 
the current experimental limit of the extra dimensions  \cite{Hoyle}. If $M_{10} \sim 100\;TeV$ we 
find that $R$ needs to be of the order of $10^{-25} \; m$. For $M_{10} \sim 100\;eV$, we 
have $R \simeq 10 \; {\mbox{microns}}$.  Therefore, brane world  of string theory on $S^{1}/Z_{2}$ 
provides  a viable mechanism to get $\rho_{\Lambda}$ down to its current observational value. 
Hence, the ADD mechanism that was initially designed to solve  the hierarchy problem \cite{ADD98} 
also solves the CC problem in string theory. 

Although the action of Eq.(\ref{2.1}) is valid only for type II and heterotic string, it is
straightforward to show that our above conclusions are also true for type I string. As a
matter of fact, the only difference will be in the expression of $T^{(D)}_{ab}$, while
all the rest  remains the same, so does Eq.(\ref{3.17}), based on which our  above conclusions were 
derived. Similarly, the addition of other matter fields, such as the Yang-Mills and Chern-Simons
terms \cite{LWC00}, in action (\ref{2.1}) does not change our conclusions either.

It is remarkable to note that the same mechanism is also valid in the framework of 
the Horava-Witten heterotic M-Theory on $S^{1}/Z_{2}$ \cite{GWW07}. All of these strongly
suggest that the above mechanism for solving the long-standing CC problem is
a built-in mechanism in the brane world  of string/M-Theory.

\section{Brane Cosmology in String Theory}

\renewcommand{\theequation}{5.\arabic{equation}}
\setcounter{equation}{0}
 
 We consider  spacetimes with  the metric \cite{WCS08}, 
\bq
\lb{4.4}
ds^{2}_{5} =  e^{2\sigma(t,y)}\left(dt^{2} - dy^{2}\right)  - e^{2\omega(t,y)}d\Sigma^{2}_{k},
\eq 
where $d\Sigma^{2}_{k} = {dr^{2}}/{(1 - k r^{2})} + r^{2}\left(d\theta^{2} 
+ \sin^{2}\theta d\phi^{2}\right)$.
Assuming that the two orbifold branes are located at   $y = y_{I}(t_{I})$,
we find that   the reduced metric takes the form,
\bq
\lb{4.4a}
\left. ds^{2}_{5}\right|_{M^{(I)}_{4}} = g^{(I)}_{\mu\nu}d\xi_{(I)}^{\mu}d\xi_{(I)}^{\nu}
= d\tau_{I}^{2} - a^{2}\left(\tau_{I}\right)d\Sigma^{2}_{k},
\eq
where $\xi^{\mu}_{(I)} \equiv \left\{\tau_{I}, r, \theta, \varphi\right\}$, and
$\tau_{I}$ denotes the proper time of the I-th brane, given by
$d\tau_{I} = e^{\sigma}\sqrt{1 - \left({\dot{y}_{I}}/{\dot{t}_{I}}\right)^{2}}\; dt_{I}$,
and $a\left(\tau_{I}\right) \equiv \exp\left\{\omega\left[t_{I}(\tau_{I}), y_{I}(\tau_{I})\right]\right\}$,
with $\dot{y}_{I} \equiv d{y}_{I}/d\tau_{I}$, etc. For the sake of simplicity and without causing any
confusion, from now on we shall drop all the indices ``I". 
The normal vector $n_{a}$ and   $e^{a}_{(\mu)}$ are given, respectively, by
$n^{a} = - \epsilon_{y} \left(\dot{y}\delta^{a}_{t} +  \dot{t}\delta^{a}_{y}\right)$, 
$e^{a}_{(\tau)} =  \dot{t}\delta^{a}_{t} +  \dot{y}\delta^{a}_{y}$,
$e^{a}_{(r)} = \delta^{a}_{r}$, $e^{a}_{(\theta)} = \delta^{a}_{\theta}$, and
$e^{a}_{(\varphi)} = \delta^{a}_{\varphi}$,
where $\epsilon_{y} = \pm 1$. Then, we find that, for a perfect fluid $\; \tau_{\mu\nu} = 
(\rho + p)u_{\mu}u_{\nu} - p g_{\mu\nu}$, where $u_{\mu} = \delta^{\tau}_{\mu}$,
the field equations on the branes are given by
\bqn
\lb{4.14a}
 H^{2} &+& \frac{k}{a^{2}} = \frac{8\pi G}{3}\left(\rho + \tau_{(\phi, \psi)}\right) 
      + \frac{1}{3}\Lambda 
      + \frac{1}{3}{\cal{G}}^{(5)}_{\tau} + E^{(5)}\nb\\
       & &  
       + \frac{2\pi G}{3\rho_{\Lambda}}\left(\rho + \tau_{(\phi, \psi)}\right)^{2},\\
\lb{4.14b}
\frac{\ddot{a}}{a}   &=&  - \frac{4\pi G}{3}\left(\rho +3p- 2\tau_{(\phi, \psi)}\right) 
       + \frac{1}{3}\Lambda \nb\\
       & &
      - \frac{1}{6}\left({\cal{G}}^{(5)}_{\tau} + 3{\cal{G}}^{(5)}_{\theta}\right)
        - E^{(5)}  
      -  \frac{2\pi G}{3\rho_{\Lambda}}\left[\rho\left(2\rho + 3p\right) \right.\nb\\
      & &  
      \left. 
      + \left(\rho + 3p - \tau_{(\phi, \psi)}\right) \tau_{(\phi, \psi)}\right],
\eqn 
where $H \equiv \dot{a}/{a}$, and  
 \bqn
 \lb{4.13}
 E^{(5)} &\equiv& \frac{1}{6}e^{-2\sigma}\left[\sigma_{,tt} - \omega_{,tt}
 - \sigma_{,yy} + \omega_{,yy} 
 + k e^{2(\sigma - \omega)}\right],\nb\\
 {\cal{G}}^{(5)}_{\tau} &\equiv& \frac{1}{3}e^{-2\sigma}\left[\left({\phi_{,t}}^{2} + {\psi_{,t}}^{2}\right)
 - \left({\phi_{,y}}^{2} + {\psi_{,y}}^{2}\right)\right]\nb\\
 & & - \frac{1}{24}\left\{5\left[\left(\nabla\phi\right)^{2}  + \left(\nabla\psi\right)^{2}\right]
 - 6 V_{5}\right\},\nb\\
 {\cal{G}}^{(5)}_{\theta} &\equiv&  \frac{1}{24}\left\{ 8 \left({\phi_{,n}}^{2} + {\psi_{,n}}^{2}\right)
 - 6 V_{5} \right.\nb\\
 & & \left. 
 + 5\left[\left(\nabla\phi\right)^{2} + \left(\nabla\psi\right)^{2}\right]\right\},
 \eqn
with $\phi_{,n} \equiv n^{a}\nabla_{a}\phi, \; \Lambda \equiv \Lambda_{4}$ and
$G \equiv G_{4}$.  The first two terms in the right-hand sides of
Eqs.(\ref{4.14a}) and (\ref{4.14b}) also appear in the Einstein's theory of gravity, although their
origins    are completely different \cite{GWW07}. The rest denotes the brane corrections,
and the effects of which on the evolution of the universe depend on specific models to be considered.

\section{Conclusions}

\renewcommand{\theequation}{6.\arabic{equation}}
\setcounter{equation}{0}

 In this Letter, we have studied orbifold branes in string theory in
(D+d)-dimensions, and obtained the general field equations both outside and 
on   the branes for type II and heterotic string.  We have  investigated the 
radion stability, using the Goldberger-Wise mechanism \cite{GW99}, and shown 
explicitly that it is stable.

We have also shown explicitly that for $D=d = 5$ the effective cosmological 
constant on  the branes can be easily lowered to its current observational value, 
using    large extra dimensions. This is also true for type I string.  
Therefore, brane world of string theory provides  a built-in 
mechanism for solving the long-standing cosmological constant problem. 

Applying the formulas to cosmology, we have obtained the generalized Friedmann equations
on each of the two branes. Investigations of their cosmological implications, including 
current acceleration of the universe, are under our current considerations. 

\section*{Acknowledgments} 

The authors thank Qiang Wu for the help of preparing the figure. Part of the work was done 
when the authors   visited LERMA/CNRS-FRE. They would like to thank the Laboratory 
for hospitality. This work was partially supported  by  NSFC under grant No. 10775119 (AW).

\end{document}